\documentclass[aps,prb,reprint,amsfonts,amsmath,amssymb,longbibliography,superscriptaddress]{revtex4-2}
\usepackage{hyperref}
\usepackage{graphicx}
\usepackage{bm}
\usepackage{newtxtext}
\usepackage[varg]{newtxmath}
\usepackage[normalem]{ulem}
\usepackage{braket}
\hypersetup{colorlinks=true,linkcolor=blue,citecolor=blue,urlcolor=blue}
\usepackage{xcolor}

\begin{document}

\title{Photocontrol of spin scalar chirality in centrosymmetric itinerant magnets}
\author{Atsushi Ono}
\affiliation{Department of Physics, Tohoku University, Sendai 980-8578, Japan}
\author{Yutaka Akagi}
\affiliation{Department of Physics, Graduate School of Science, The University of Tokyo, 7-3-1 Hongo, Tokyo 113-0033, Japan}

\begin{abstract}
Noncoplanar magnetic structures, such as magnetic skyrmions, are characterized by spin chirality and usually favored by antisymmetric exchange interactions in noncentrosymmetric magnets.
Here, we show that a linearly polarized electric-field pulse stabilizes a nonequilibrium spin scalar chiral state in a centrosymmetric itinerant ferromagnet.
The scalar chirality has a nonmonotonic dependence on the electric-field strength, and its sign can be controlled by circular polarization.
Furthermore, magnetic skyrmions are excited after the pulse decays.
A photoinduced nonthermal electron distribution plays an important role for instability towards the spin scalar chiral state as well as the $120^{\circ}$ N\'eel state, depending on the next-nearest-neighbor transfer integral.
These results provide an alternative route to controlling spin chirality by photoirradiation.
\end{abstract}

\maketitle

\textit{Introduction.}
Swirling spin textures, e.g., magnetic bubbles, vortices, and skyrmions, have been intensively studied in condensed matter physics~\cite{Seki2016,Han2017,Zang2018,Blachowicz2019,Nagaosa2013,Ochoa2018,Back2020,Fujishiro2020,Tokura2021,Batista2016,Hayami2021}.
These noncoplanar spin orders are usually attributed to the competition between the ferromagnetic (FM) interaction and the antisymmetric Dzyaloshinskii--Moriya (DM) interaction, the latter of which is present in noncentrosymmetric systems.
Recently, the skyrmion crystal was discovered even in centrosymmetric magnets without the DM interaction~\cite{Kurumaji2019,Hirschberger2019,Khanh2020,Yasui2020}, where, e.g., geometrical frustration, magnetic anisotropy, and Fermi-surface instability play a role in generating spin chirality~\cite{Okubo2012,Leonov2015,Amoroso2020,Martin2008,Akagi2010,Akagi2012,Ozawa2017,Hayami2017a,Kumar2010}.
When electrons are subject to the emergent fields associated with noncoplanar or noncollinear spin structures, intriguing transport~\cite{Lee2009,Neubauer2009,Shiomi2013,Hirschberger2020,Kim2019s,Ishizuka2020a,Fujishiro2021} and optical~\cite{Hayashi2021,Sorn2021,Feng2020k,Okumura2021,Mochizuki2010b} phenomena are observed, thereby attracting scientific and technological interest.

In the past decade, it has been demonstrated that optical pulses enable the ultrafast control of spin chirality.
In multiferroics, the magnetic structure can be switched by terahertz pulses or x-ray irradiation~\cite{Mochizuki2010,Kubacka2014,Yamasaki2015,Sato2016,Khan2020}.
Also, many researchers have studied the optical creation of skyrmions in noncentrosymmetric magnets, where the local excitation by a laser promotes the nucleation of the skyrmions~\cite{Finazzi2013,Koshibae2014,Ogawa2015,Heo2016,Flovik2017,Berruto2018,Fujita2017,Je2018,Yang2018}.
The nonthermal control of magnetic interactions and spin chirality via a high-frequency laser, microwave, and static fields was also proposed theoretically~\cite{DelaTorre2021,Sato2016,Bukov2016,Losada2019,Taguchi2012,Claassen2017,Kitamura2017,Eto2021,Owerre2017,VinasBostrom2020,Stepanov2017,VinasBostrom2022,Quito2021,Kobayashi2021,Inoue2022,Ghosh2022,Yudin2017,Hanslin2021,Furuya2021}.
For example, circularly polarized light breaks the time-reversal symmetry and can be coupled to the scalar chirality~\cite{Taguchi2012,Claassen2017,Kitamura2017,Eto2021,Owerre2017,VinasBostrom2020}.
Since linearly polarized light and a static electric field break inversion symmetry, they can modulate or induce the DM interaction and magnetic anisotropy~\cite{Yudin2017,Hanslin2021,Furuya2021}.
However, these studies have considered noncentrosymmetric systems where the DM interaction is present or dynamically induced through spin-orbit coupling, except for Refs.~\cite{Inoue2022,Ghosh2022} discussing spin vector chirality.

\begin{figure}[b]\centering
\includegraphics[scale=1]{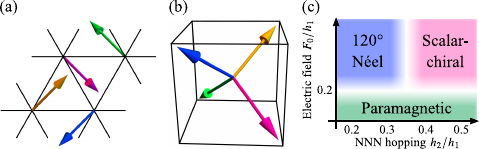}
\caption{(a)~Snapshot of spin vectors in a four-sublattice spin scalar chiral state.
(b)~The four spin vectors point to the four vertices of a regular tetrahedron.
(c)~Schematic steady-state phase diagram.}
\label{fig:1}
\end{figure}

In this Letter, we investigate the real-time dynamics induced by an electric field in a centrosymmetric itinerant ferromagnet described by the double-exchange model on a triangular lattice.
A detailed ground-state phase diagram of this model was shown in Ref.~\cite{Akagi2010}; a four-sublattice spin scalar chiral state [see Figs.~\ref{fig:1}(a) and \ref{fig:1}(b)] is present near $1/4$ and $3/4$ fillings, exhibiting an anomalous Hall effect.
First, we show that a static electric field can stabilize either the spin scalar chiral state or the $120^\circ$ N\'eel state on a picosecond timescale and obtain a schematic ``steady-state phase diagram'' shown in Fig.~\ref{fig:1}(c).
A nonthermal electron distribution due to irradiation induces the instability to these nonequilibrium steady states.
Then, we demonstrate that the sign of the scalar chirality can be selected by circular polarization.
The steady states with scalar chirality are induced by a terahertz pulse as well.
Finally, we show that metastable magnetic skyrmions emerge after the pulse irradiation.

\begin{figure*}[t]\centering
\includegraphics[scale=1]{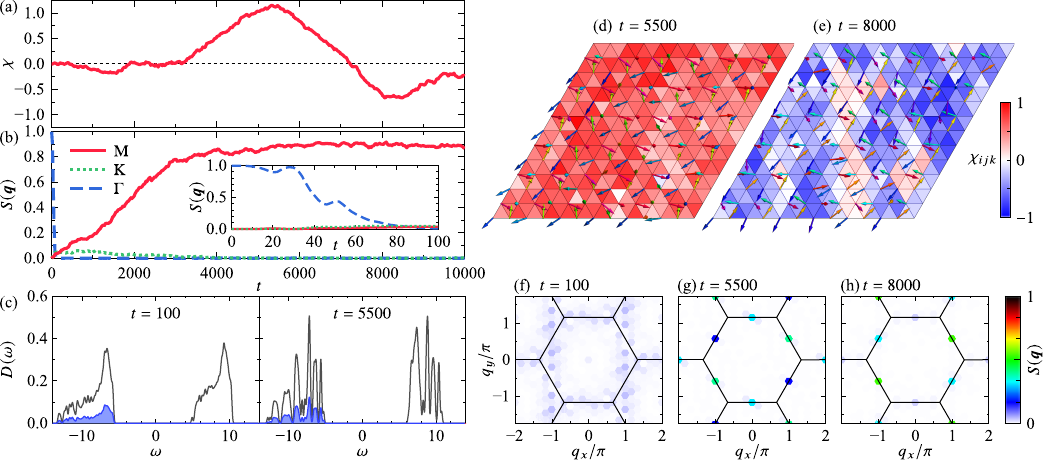}
\caption{(a),(b)~Time profiles of (a)~spin scalar chirality and (b)~the spin structure factor with a static electric field.
The inset in (b) is an enlarged view.
(c)~Density of states $D(\omega) = \sum_\nu \delta(\omega-\varepsilon_\nu)$ (gray solid lines) and the occupation $\sum_\nu n_\nu \delta(\omega-\varepsilon_\nu)$ (blue shades).
(d), (e)~Snapshots of the localized spins and local spin chirality.
The arrow colors indicate the in-plane angles.
(f)--(h)~Spin structure factor in reciprocal space.
The solid lines indicate the Brillouin zone boundary.
NNN transfer integral is set to $h_2 = 0.4$ in (a)--(h).}
\label{fig:2}
\end{figure*}

\textit{Model and method.}
We consider the FM Kondo lattice model on a triangular lattice defined by the Hamiltonian
\begin{align}
\mathcal{H} = \sum_{ij s} h_{ij} c_{is}^\dagger c_{js} - J \sum_{iss'} c_{is}^\dagger \bm{\sigma}_{ss'} c_{is'} \cdot \bm{S}_i.
\label{eq:Ham}
\end{align}
Here, $c_{is}^\dagger$ ($c_{is}$) denotes the creation (annihilation) operator of an itinerant electron with spin $s\ (= {\uparrow}, {\downarrow})$ at site $i$, and the vectors $\bm{\sigma}$ and $\bm{S}_i$ represent the Pauli matrix and a classical spin localized at site $i$ with $|\bm{S}_i|=1$, respectively.
The coefficients $h_{ij}$ and $J$ denote the transfer integral and exchange interaction, respectively.
We assume that $h_{ij} = -h_1$ for the nearest neighbor (NN), $h_{ij} = -h_2$ for the next nearest neighbor (NNN), and $h_{ij} = 0$ for the rest, and adopt a finite cluster of $N=12\times 12$ sites with periodic boundary conditions.
The parameters are set to $J/h_1 = 8$, $h_2/h_1 = 0.2$--$0.5$, and $N_{\mathrm{e}}/N = 1/4$ with $N_{\mathrm{e}}$ being the number of electrons (i.e., $1/8$ filling) so that the ground state is a FM metallic state.
Hereafter, the elementary charge $e$, Dirac constant $\hbar$, lattice constant $a$, and NN transfer integral $h_1$ are set to unity; energy, time, and electric-field strength are expressed in units of $h_1$, $\hbar/h_1 = 0.66\ \mathrm{fs}$, and $h_1/(ea) = 20\ \mathrm{MV/cm}$ for $h_1 = 1\ \mathrm{eV}$ and $a = 5\ \mathrm{\mathring{A}}$, respectively.
An external vector potential $\bm{A}$ is incorporated via the Peierls substitution $h_{ij} \to h_{ij} e^{i\bm{A}(t)(\bm{r}_i-\bm{r}_j)}$, where $t$ denotes time and $\bm{r}_i$ is the position of site $i$.
The time evolution of the electrons and spins is governed by the von Neumann equation $\dot{\rho} = -i [\mathcal{H}, \rho]$ and the Landau--Lifshitz--Gilbert (LLG) equation $\dot{\bm{S}}_i = \bm{S}_i \times \bm{b}_i - \alpha \bm{S}_i \times \dot{\bm{S}}_i$, respectively~\cite{Koshibae2009,Chern2018,Ono2017,Ono2018,Ono2020}.
Here, $\rho_{is,js'} = \langle c_{js'}^\dagger c_{is} \rangle$ is the one-body density matrix of the electrons, $\bm{b}_i$ represents an effective field given by $\bm{b}_i = -\partial \langle \mathcal{H} \rangle / \partial \bm{S}_i$, and $\alpha$ denotes the Gilbert damping constant.
These equations are solved by the fourth-order Runge--Kutta method with a time step of $\delta t = 0.01$.
Even though dissipation is introduced only in the LLG equation, the electrons are relaxed and scattered via the coupling term~\cite{Koshibae2009}.
By diagonalizing the Hamiltonian at time $t$, we can obtain the single-particle energy level $\varepsilon_\nu(t)$ and its occupation number $n_\nu(t)$.
The damping constant is set to $\alpha = 1$.
We introduce a random tilt from the FM ground state up to $\delta\theta_{\mathrm{max}} = 0.1$ rad into $\{\bm{S}_i\}$ at $t = 0$, which mimics thermal fluctuations~\cite{Koshibae2009}; we need to consider the average over the initial configurations of $\{\bm{S}_i\}$ when discussing the macroscopic spin-texture dynamics.

To investigate the dynamics of the magnetic structure, we consider the spin scalar chirality $\chi = N^{-1} \sum_{i} (\chi_{\bm{r}_i,\bm{r}_i+\bm{a}_1,\bm{r}_i+\bm{a}_2} + \chi_{\bm{r}_i+\bm{a}_1+\bm{a}_2,\bm{r}_i+\bm{a}_2,\bm{r}_i+\bm{a}_1})$, where $\bm{a}_1 = (1,0)$, $\bm{a}_2 = (1/2,\sqrt{3}/2)$, and $\chi_{\bm{r}_i,\bm{r}_j,\bm{r}_k} = \chi_{ijk} = \bm{S}_i \cdot (\bm{S}_j \times \bm{S}_k)$.
We also compute the spin structure factor defined by $S(\bm{q}) = N^{-2} \sum_{ij} \bm{S}_i \cdot \bm{S}_j e^{i\bm{q} \cdot (\bm{r}_i - \bm{r}_j)}$, which satisfies $S(-\bm{q}) = S(\bm{q})^* = S(\bm{q})$ and $\sum_{\bm{q}} S(\bm{q}) = 1$.
Considering that $S(\bm{q})$ has a peak at the $\mathrm{M}$ ($\mathrm{K}$) point in the spin scalar chiral ($120^\circ$ N\'eel) state, we define $S(\mathrm{M}) = S(\pi,-\pi/\sqrt{3}) + S(\pi,\pi/\sqrt{3}) + S(0,2\pi/\sqrt{3})$ and $S(\mathrm{K}) = S(4\pi/3,0) + S(2\pi/3, 2\pi/\sqrt{3})$.
Note that $\vert\chi\vert = 8/(3\sqrt{3}) \approx 1.54$ and $S(\mathrm{M}) = 1$ in the spin scalar chiral state~\cite{Martin2008, Akagi2010}, and $\chi = 0$ and $S(\mathrm{K}) = 1$ in the $120^\circ$ N\'eel state.

\textit{Results.}
We show the real-time dynamics in Fig.~\ref{fig:2} when we apply a linearly polarized static electric field $\bm{F}(t) = -\dot{\bm{A}}(t) = (F_0, 0)$ with $F_0 = 0.5$, which precludes conventional mechanisms for inducing scalar chirality.
See Supplemental Material~\footnote{See Supplemental Material at \href{http://link.aps.org/supplemental/10.1103/PhysRevB.108.L100407}{http://link.aps.org/supplemental/ 10.1103/PhysRevB.108.L100407} for a video of the time evolution of the spin structure, and for detailed discussions on the winding number, short-range correlation of scalar chirality, and early-time dynamics.} for a video of the spin configuration, spin scalar chirality, and structure factor.
The dynamics can be divided into the following two stages.
(i)~The collapse of the FM order ($t \lesssim 100$):
After the electric field is switched on at $t = 0$, the FM correlation $S(\Gamma) = S(0,0)$ rapidly decays at $t \approx 40$ as shown in Fig.~\ref{fig:2}(b).
At $t = 100$, the electrons uniformly occupy the lower band [the left panel of Fig.~\ref{fig:2}(c)], and $S(\bm{q})$ is finite near the Brillouin zone boundary [Fig.~\ref{fig:2}(f)].
The instability of the FM order is induced by the inverted electron distribution in the lower band due to the Bloch oscillation of the electrons~\cite{Ono2020, Note1}.
Here, the Bloch-oscillation period is given by $4\pi/F_0 \approx 25.1$, which is reflected in $S(\Gamma)$ shown in Fig.~\ref{fig:2}(b).
(ii)~The development of the spin scalar chirality ($t \gtrsim 100$):
The scalar-chiral correlation $S(\mathrm{M})$ builds up linearly in time until $t \approx 3000$.
The scalar chirality $\chi$ starts to increase at $t \approx 3000$, and then takes the maximal value $\approx 1.1$ at $t = 5500$, as displayed in Fig.~\ref{fig:2}(a).
The uniform spin scalar chiral structure is seen in Fig.~\ref{fig:2}(d).
After $t \approx 5500$, the scalar chirality $\chi$ decreases down to $\approx -0.7$ while $S(\mathrm{M})$ remains constant.
In Fig.~\ref{fig:2}(e), we observe the spin scalar chiral state with $\chi < 0$ at $t = 8000$.
For $t \gtrsim 3000$, the structure factor has three peaks at $\bm{q} = (\pi,-\pi/\sqrt{3})$, $(\pi,\pi/\sqrt{3})$, and $(0,2\pi/\sqrt{3})$ as shown in Figs.~\ref{fig:2}(g) and \ref{fig:2}(h), although their values slightly deviate from $1/3$.
The sign of $\chi$ is unstable in time in the static field; we will later show that circularly polarized light can control and stabilize the sign.

Although the spin scalar chirality is not completely uniform as shown in Figs.~\ref{fig:2}(d) and \ref{fig:2}(e), we can confirm the topological nature by calculating the Chern number $n_\mathrm{Ch}$.
In aperiodic systems, such index is defined as~\cite{Avron1994a, Avron1994b, Aizenman1998} \footnote{We have other expressions of the Chern number in aperiodic systems, such as the Bott index (see, e.g., Refs.~\cite{Bellissard1994, Hastings2010, Loring2010}) and the Chern number in the real space~\cite{Kitaev2006, Prodan2010}.
The nonzero index corresponds to the system showing the topological Hall effect.}
\begin{align}
n_\mathrm{Ch} \coloneqq \mathrm{dim} \ker [A - 1] - \dim \ker [A + 1], 
\label{eq:index}
\end{align}
where $A\coloneqq P_\mathrm{F} - \mathcal{D}_{\bm a}^{\dagger} P_\mathrm{F} \mathcal{D}_{\bm a}$, 
$P_\mathrm{F} = \sum_{\varepsilon_\nu \le \varepsilon_\mathrm{F}} \ket{\nu}\bra{\nu}$, 
and $\mathcal{D}_{\bm a}({\bm x}) \coloneqq {\left[x + i y - (a_x + i a_y)\right]}/{\left| x + i y - (a_x + i a_y) \right|}$.
Here, $\ket{\nu}$ are eigenstates of the Hamiltonian (\ref{eq:Ham}),
${\bm x} =$ \mbox{$(x, y)$} is the position vector of the lattice points, 
and ${\bm a} =$ \mbox{$(a_x, a_y) \neq {\bm x} $} is a vector of a flux point (see, e.g., Refs.~\cite{Akagi2017, Akagi2020} for the detailed calculation).
From Eq.~(\ref{eq:index}), we obtain $n_\mathrm{Ch}=1$ and $-1$ for the lowest band at $t=5500$ [Fig.~\ref{fig:2}(d)] and $t=8000$ [Fig.~\ref{fig:2}(e)], respectively,
indicating that these spin scalar chiral states have a topological nature.
However, the dc Hall conductivity would be canceled out in the photoinduced spin scalar chiral state since the total Chern number is zero for the effective half-filled case where the lower two bands are uniformly occupied, while we might see the Hall current or edge state during the decay process of itinerant electrons after switching off photoirradiation.
Nevertheless, the photoinduced spin scalar chiral state can exhibit optical Hall effects owing to the nonzero Chern number
because optical interband transitions are still allowed.
The calculation of physical quantities for itinerant electrons reflecting the topological nature, such as optical Hall conductivity, in real-time evolution will be reported in a future paper.

The appearance of the spin scalar chiral state is ascribed to that uniform electron distribution in the lower band which minimizes the total energy of the spin scalar chiral state rather than the FM state.
As shown in Fig.~\ref{fig:2}(c), the electrons uniformly occupy the lower band(s) after the collapse of the FM order, implying that the energy bands are effectively half filled.
It was shown that the ground state of the present model at half filling is either $120^\circ$ N\'eel or spin scalar chiral state depending on the NNN transfer integral $h_{2}$ \footnote{The details will be reported elsewhere.}, whereas we expect that the high-temperature state is paramagnetic, given the finite-temperature phase diagrams~\cite{Kumar2010, Kathyat2020}.
Our results are roughly consistent with ground-state phase diagrams, suggesting that this spin scalar chiral state is not due to thermal excitations.

\begin{figure}[t]\centering
\includegraphics[scale=1]{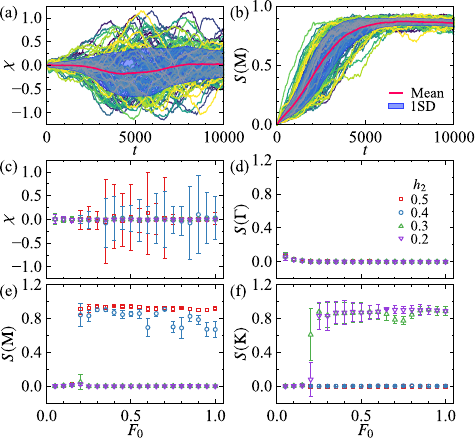}
\caption{Time profiles of (a)~$\chi$ and (b)~$S(\mathrm{M})$ with $h_2 = 0.4$.
The red line and blue shade denote the mean and standard deviation at each time, respectively.
(c)--(f)~Mean and standard deviation at $t = 10\, 000$ for $h_2 = 0.2$--$0.5$.
}
\label{fig:3}
\end{figure}

Next, we investigate the random tilt introduced into $\{\bm{S}_i\}$ at $t = 0$.
Figures~\ref{fig:3}(a) and \ref{fig:3}(b) show the time profiles of $\chi$ and $S(\mathrm{M})$ for $100$ different initial configurations of $\{\bm{S}_i\}$.
The scalar chirality appears for $t \gtrsim 2000$ and varies between $-1.2$ and $1.2$, while the absolute value of the mean, plotted as the red curve, is always less than $0.2$.
This indicates that the sign of $\chi$ depends on the initial configurations and is not controlled by the static field, and thus, the present spin scalar chiral state is not due to the field-induced DM interactions.
Meanwhile, $S(\mathrm{M})$, which reflects the long-range correlation of the scalar chirality, gradually builds up and is kept above $0.8$ for $t \gtrsim 5000$ even when $\chi \approx 0$.
We also found that the short-range correlation of scalar chirality is stronger than that in disordered states~\cite{Note1}.
Therefore, we consider the spin scalar chiral state to be stable in the thermodynamic limit, although the size of the domain with positive or negative chirality varies with time.

In Figs.~\ref{fig:3}(c)--\ref{fig:3}(f), we show the mean and standard deviation of the chirality and structure factor at $t = 10\, 000$.
The spin scalar chiral or $120^\circ$ N\'eel state is stabilized for $F_0 \gtrsim 0.2$, depending on $h_2$.
From these data, we sketch a ``steady-state phase diagram'' in Fig.~\ref{fig:1}(c), where there are three phases, i.e., the spin scalar chiral phase for $F_0 \gtrsim 0.2$ and $h_2 \gtrsim 0.4$, the $120^\circ$ N\'eel phase for $F_0 \gtrsim 0.2$ and $h_2 \lesssim 0.3$, and the paramagnetic phase for $0 < F_0 \lesssim 0.2$.

\begin{figure}[t]\centering
\includegraphics[scale=1]{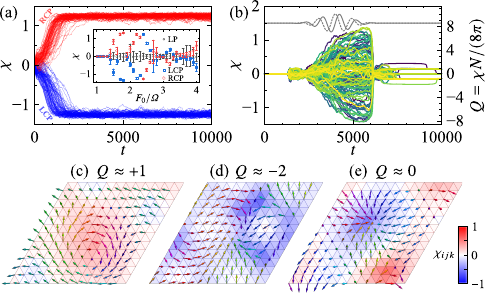}
\caption{(a)~Time profiles of $\chi$ with the RCP (red lines) and LCP (blue lines) light with $F_0 = 2.2$ and $\varOmega = 1$.
The inset shows the mean and standard deviation at $t = 10\, 000$ as a function of $F_0/\varOmega$.
(b)~Time profiles of $\chi$ with an RCP pulse.
The electric field of the $x$ ($y$) component is plotted as the gray solid (dashed) line.
Parameters are set to $F_0 = 1.8$, $\varOmega = 2\pi/1000$, and $\delta\theta_{\mathrm{max}} = 1$.
(c)--(e)~Snapshots of the spin structure and local scalar chirality at $t = 8000$.
NNN transfer integral is set to $h_2 = 0.5$ in (a)--(e).}
\label{fig:4}
\end{figure}

The sign of the scalar chirality can be controlled by circular polarization.
Figure~\ref{fig:4}(a) shows the time profiles of $\chi$ with left/right circularly polarized (LCP/RCP) light defined by $\bm{A}(t) = (F_0/\varOmega)(\pm \cos \varOmega t , \sin \varOmega t )$ with $F_0 = 2.2$ and $\varOmega = 1$, for $100$ initial configurations.
In the steady states, $\chi \approx 1.2$ for RCP and $\chi \approx -1.2$ for LCP light.
The mean and standard deviation at $t = 10\, 000$ are plotted in the inset of Fig.~\ref{fig:4}(a), as a function of $F_0/\varOmega$.
The spin scalar chiral state is stabilized for $F_0/\varOmega \gtrsim 1.4$, and the mean values of $\chi$ exhibit a nonmonotonic behavior with respect to $F_0/\varOmega$, which is beyond the perturbation theory for the coupling between scalar chirality and circular polarization~\cite{Taguchi2012}.
Furthermore, the signs of $\chi$ for LCP and RCP light are opposite to each other, while the mean values of $\chi$ for linearly polarized (LP) light $\bm{A}(t) = (F_0/\varOmega)(\sin\varOmega t,0)$ are almost zero.
Note that the controllability of the sign of $\chi$ is observed for $\varOmega = 1$ but not for $\varOmega = 20$ and $40$ (not shown); this may imply that the light frequency should be comparable to the transfer integral $h_1$ and much smaller than the band gap ($\sim 2J$).
A detailed analysis for the circular polarization is left for future work.

Lastly, we demonstrate the real-time dynamics induced by an RCP pulse that contains a few optical cycles, $\bm{A}(t) = (F_0/\varOmega) \exp[-(t-t_0)^2/(2t_{\mathrm{w}}^2)] (\cos [\varOmega (t-t_0)], -\sin [\varOmega (t-t_0)])$ with $t_0 = 4000$ and $t_{\mathrm{w}} = 800$.
In Fig.~\ref{fig:4}(b), we observe that the scalar chirality develops up to $\pm 1.2$ during irradiation and rapidly decays after irradiation.
However, the sign of $\chi$ is not controlled by either a RCP or LCP pulse since the central frequency of the pulse, $\varOmega = 2\pi/1000 \approx 0.0063 \ll h_1$, is quite low, so that the dynamics are close to those in the static field.
Here, the advantage of the circularly polarized pulse is that the magnitude of the electric field does not vanish during irradiation in contrast to the LP pulse, which prevents the relaxation to the FM ground state and promotes the growth of $\chi$.

After pulse irradiation, a metastable state with approximate integer $Q = \chi N/(2\times 4\pi)$ is realized, and it survives until $t \sim 10\, 000$.
Here, $Q$ is the Pontryagin number~\cite{Note1,Lohani2019}, which counts the net number of skyrmions in the cluster.
At $t = 8000$, a skyrmionlike magnetic object is observed in Figs.~\ref{fig:4}(c)--\ref{fig:4}(e).
When a single skyrmion is present as shown in Fig.~\ref{fig:4}(c), $Q$ is approximately equal to $\pm 1$.
In the present calculations for $N = 12\times 12$, we also observe states with $|Q| \approx 2$ as displayed in Fig.~\ref{fig:4}(d).
Figure~\ref{fig:4}(e) exemplifies a state in which both $Q \approx 1$ and $Q \approx -1$ objects are created while the total $Q$ is canceled out.
The lifetime of the metastable states depends on the relaxation rate, i.e., the Gilbert damping constant $\alpha$ in the present calculations; we found that the lifetime increases with decreasing $\alpha$ as expected.
Note that the metastable skyrmions can be excited by the LP pulse as well and even without $h_2$.

\textit{Discussion.}
The required magnitude of the electric field is estimated to be of the order of $\hbar/(ea\tau) \sim 1\ \mathrm{MV/cm}$ with $\tau$ being an intraband relaxation time~\cite{Ono2020} (typically, $\tau \gtrsim 10\ \mathrm{fs}$).
This magnitude is so strong that materials will be damaged by the static field.
However, this problem can be circumvented by using a short terahertz pulse~\cite{Fulop2020}, as demonstrated in Fig.~\ref{fig:4}.

The time scale of the dynamics is determined by the Gilbert damping constant $\alpha$ as well as the field magnitude.
In real materials in equilibrium, $\alpha$ is of order $0.001$--$0.01$~\cite{Maekawa2017}, which is two or three orders of magnitude smaller than $\alpha = 1$ adopted here.
We found that, for $\alpha = 0.01$, the long-range spin scalar chiral or $120^\circ$ N\'eel order is suppressed in the steady states.
If a large $\alpha$ of order $1$ is essential, the realization of these photoinduced orders would be difficult.
Theoretically, we might overcome this difficulty by introducing another magnetic interaction that favors the spin scalar chiral state, e.g., an antiferromagnetic (AFM) interaction between the localized spins~\cite{Akagi2010,Akagi2011,Akagi2015}, or by considering the dissipation due to electron-electron interactions far from equilibrium.

\textit{Summary and outlook.}
We have investigated the real-time dynamics in a centrosymmetric itinerant magnet and found that the spin scalar chiral or $120^\circ$ N\'eel state is stabilized by a static field, continuous-wave light, and low-frequency pulse, depending on the NNN transfer integral.
The sign of scalar chirality can be selected by circular polarization.
Additionally, skyrmions are created in metastable states after pulse irradiation, characterized by the Pontryagin number.
These spin scalar chiral states and metastable skyrmion states arise from the nonthermal electron distribution, not from antisymmetric interactions such as the DM interaction.
The underlying mechanism should be further clarified in future work.

Recently, it was revealed that the AFM skyrmionlike textures are transiently created on the centrosymmetric square lattice~\cite{Ono2019}.
In this study, however, the spin scalar chiral state is stable under optical fields and controllable by circular polarization.
Furthermore, contrary to the conventional theory~\cite{Taguchi2012}, our results demonstrate that neither the DM interaction nor circular polarization is necessary for the spin scalar chiral state; the absence of antisymmetric interactions makes the optical switching of scalar chirality easier.
These findings suggest that centrosymmetric magnets are promising for ultrafast control of scalar chirality.

\textit{Acknowledgments.}
This work was supported by JSPS KAKENHI Grants No.\ JP18H05208, No.\ JP19K23419, No.\ JP20K14394, No.\ JP23K13052, No.\ JP23H01108, No.\ JP20K14411, and JSPS Grant-in-Aid for Scientific Research on Innovative Areas ``Quantum Liquid Crystals'' (KAKENHI Grants No.\ JP20H05154 and No.\ JP22H04469). 
Y.A. is supported by JST PRESTO Grant No.\ JPMJPR2251.
The numerical calculations were performed using the facilities of the Supercomputer Center, the Institute for Solid State Physics, the University of Tokyo.
The authors acknowledge the fruitful discussions at the 9th workshop on ``Frontier of Theory for Condensed Matter Systems'' in Hokkaido in 2020, which shaped the early stages of this research.

\bibliography{reference}

\end{document}


\title{Supplemental Material for\\ ``Photocontrol of spin scalar chirality in centrosymmetric itinerant magnets''}
\author{Atsushi Ono}
\affiliation{Department of Physics, Tohoku University, Sendai 980-8578, Japan}
\author{Yutaka Akagi}
\affiliation{Department of Physics, Graduate School of Science, The University of Tokyo, 7-3-1 Hongo, Tokyo 113-0033, Japan}

\maketitle

\section*{Winding number}
\begin{figure}[b]\centering
\includegraphics[scale=1]{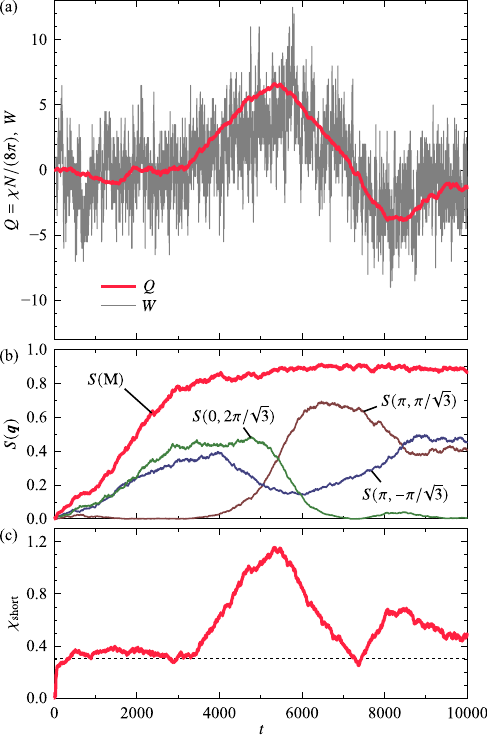}
\caption{Time profiles of (a)~scalar chirality and winding number, (b)~the spin structure factor, and (c)~the short-range correlation of the scalar chirality.
The parameters are the same as in Fig.~2.}
\label{fig:s1}
\end{figure}

\begin{figure}[b]\centering
\includegraphics[scale=1]{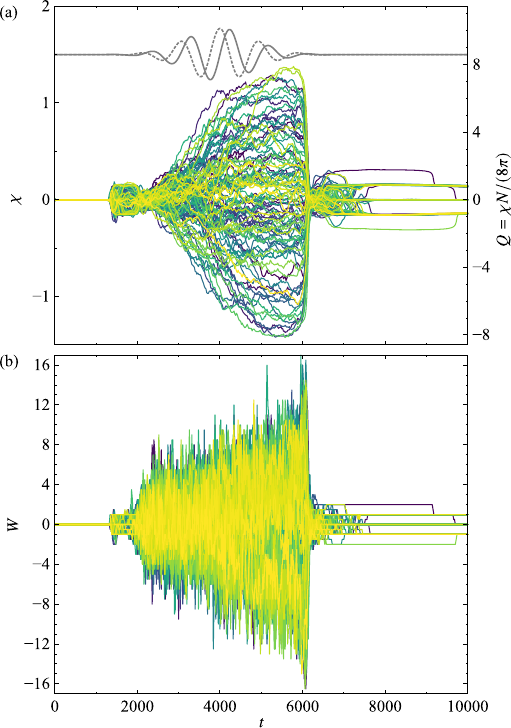}
\caption{Time profiles of (a)~scalar chirality and (b)~the winding number for $100$ initial configurations.
The parameters are the same as in Fig.~4(b).}
\label{fig:s2}
\end{figure}

According to Ref.~\cite{Lohani2019}, the winding number of spin textures defined on a triangular lattice is given by
\begin{align}
W = \frac{1}{4\pi} \sum_{i} (\mathcal{A}_{\bm{r}_i,\bm{r}_i+\bm{a}_1,\bm{r}_i+\bm{a}_2} + \mathcal{A}_{\bm{r}_i+\bm{a}_1+\bm{a}_2,\bm{r}_i+\bm{a}_2,\bm{r}_i+\bm{a}_1}), \label{eq:winding}
\end{align}
where $\mathcal{A}_{\bm{r}_i, \bm{r}_j, \bm{r}_k} = \mathcal{A}_{ijk}$ is the solid angle subtended by three spin vectors, $\bm{S}_i$, $\bm{S}_j$, and $\bm{S}_k$:
\begin{align}
\mathcal{A}_{ijk} = 2 \tan^{-1} \frac{\bm{S}_i \cdot (\bm{S}_{j} \times \bm{S}_{k})}{1 + \bm{S}_i \cdot \bm{S}_{j} + \bm{S}_j \cdot \bm{S}_{k} + \bm{S}_{k} \cdot \bm{S}_{i}}.
\end{align}
The solid angle $\mathcal{A}_{ijk}$ can be approximately given by $\mathcal{A}_{ijk} \approx \chi_{ijk}/2$ with $\chi_{ijk} = \bm{S}_{i}\cdot(\bm{S}_j\times \bm{S}_k)$ when $\mathcal{A}_{ijk}$ is small; and thus, we used $Q = \chi N/(8\pi)$ in the text.

Figure~\ref{fig:s1}(a) shows the time profiles of $Q$ and $W$ when the static electric field is applied.
The case of the circularly polarized pulse is also shown in Fig.~\ref{fig:s2}.
Both $Q$ and $\chi$ follow the same trends, while $W$ defined in Eq.~\eqref{eq:winding} is always quantized.
Therefore, the scalar chirality $\chi$ is suitable for characterizing the spin scalar chiral state.
After the pulse irradiation, the winding number is an integer from $-2$ to $+2$ [Fig.~\ref{fig:s2}(b)], indicating that the metastable states are, in fact, skyrmion states.

\section*{Short-range correlation of scalar chirality}
\begin{figure}[t]\centering
\includegraphics[scale=1]{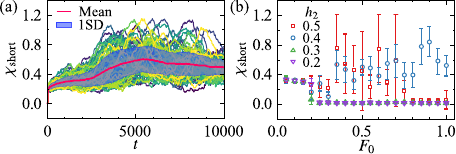}
\caption{(a)~Time profiles of $\chi_{\mathrm{short}}$ and (b)~mean values of $\chi_{\mathrm{short}}$ at $t=10\, 000$ for $100$ initial configurations.
The static electric field is applied.
The parameters are the same as in Fig.~3.}
\label{fig:s3}
\end{figure}

Here, we consider the short-range correlation of the scalar chirality,
\begin{align}
\chi_{\mathrm{short}}
&= \frac{1}{3N} \sum_{i} \left\vert \hspace{1mm} \includegraphics[valign=c,scale=0.4]{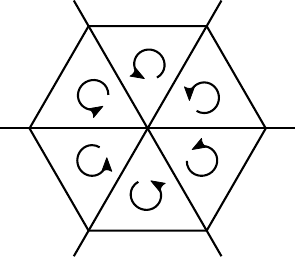} \hspace{1mm} \right\vert \notag\\
&= \frac{1}{3N} \sum_{i} \vert
\chi_{\bm{r}_i,\bm{r}_i+\bm{a}_1,\bm{r}_i+\bm{a}_2}
+ \chi_{\bm{r}_i,\bm{r}_i+\bm{a}_2,\bm{r}_i-\bm{a}_1+\bm{a}_2} \notag \\
&\quad + \chi_{\bm{r}_i,\bm{r}_i-\bm{a}_1+\bm{a}_2,\bm{r}_i-\bm{a}_1}
+ \chi_{\bm{r}_i,\bm{r}_i-\bm{a}_1,\bm{r}_i-\bm{a}_2} \notag \\
&\quad + \chi_{\bm{r}_i,\bm{r}_i-\bm{a}_2,\bm{r}_i+\bm{a}_1-\bm{a}_2}
+ \chi_{\bm{r}_i,\bm{r}_i+\bm{a}_1-\bm{a}_2,\bm{r}_i+\bm{a}_1}
\vert.
\end{align}
Note that $\chi_{\mathrm{short}} = \vert\chi \vert = 8/(3\sqrt{3}) \approx 1.54$ if the magnetic structure is uniform spin scalar chiral, and $\chi_{\mathrm{short}} \approx 0.303$ if it is paramagnetic.

Figures~\ref{fig:s1}(b) and \ref{fig:s1}(c) show the time evolution of $S(\bm{q})$ and $\chi_{\mathrm{short}}$ in the static electric field.
The uniform scalar chirality $Q \propto \chi$ slowly fluctuates in time after $S(\mathrm{M})$ approaches a constant value $\approx 0.9$, although $S(\bm{q})$ at $\bm{q} = (0,2\pi/\sqrt{3})$, $S(\pi, \pi/\sqrt{3})$, and $S(\pi, -\pi/\sqrt{3})$ are inequivalent since the rotational symmetry is broken in nonequilibrium states.
Meanwhile, for $t\gtrsim 3500$, the short-range correlation $\chi_{\mathrm{short}}$ is larger than $0.303$ [indicated by the dotted line in Fig.~\ref{fig:s1}(c)]; this can also be seen in Fig.~\ref{fig:s3}, where the time evolution of $\chi_{\mathrm{short}}$ and its average at $t=10\, 000$ are shown for $100$ initial configurations.
These observations support the conclusion that, after $S(\mathrm{M})$ becomes constant, the present spin scalar chiral state is ``stable'' in the thermodynamic limit, while the size of the domain with positive or negative chirality varies with time (see also the supplemental video).

\section*{Early-time dynamics: Bloch oscillation}
\begin{figure}[t]\centering
\includegraphics[scale=1]{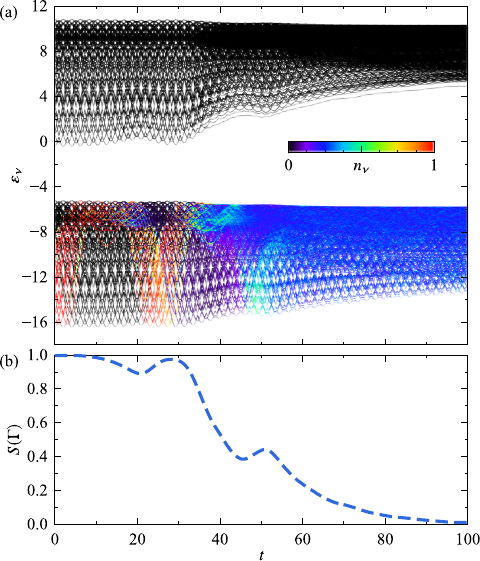}
\caption{Time profiles of (a)~the single-particle energy level $\varepsilon_\nu$ and its occupation number $n_\nu$, and (b)~the spin structure factor at the $\Gamma$ point, $S(\Gamma)$.
The parameters are the same as in Fig.~2.
}
\label{fig:s4}
\end{figure}

Here, we demonstrate that early-time dynamics are governed by the Bloch oscillation of the itinerant electrons.
Figure~\ref{fig:s4} shows the $\nu$th single-particle energy levels $\varepsilon_\nu$ and its occupation number $n_\nu$.
The ferromagnetic structure factor $S(\Gamma)$ is also plotted in Fig.~\ref{fig:s4}(b).
We find that $S(\Gamma)$ decreases when the electrons occupy the higher levels in the lower band, and the localized spins recover the ferromagnetic correlation when the electrons occupy the lower levels.
This oscillation of the electron occupation driven by the electric field, called the Bloch oscillation, is due to the periodicity of the Brillouin zone.
The time period of the Bloch oscillation is inversely proportional to the lattice constant $a$ and the electric-field magnitude $F_0$, and the oscillation period should be shorter than the relaxation time of the electrons $\tau$.
Therefore, the electric-field magnitude required for the present magnetic transition is estimated to be of the order of $\hbar/(ea\tau) \sim 1\ \mathrm{MV/cm}$ for $\tau = 10\ \mathrm{fs}$ and $a = 0.5\ \mathrm{nm}$.

\bibliography{reference}